\documentclass[a4paper,11pt]{article}
\usepackage{jheppub} 
\usepackage{lineno}
\usepackage{cleveref}

\title{Exotic Field Theories for (Hybrid) Fracton Phases from Imposing Constraints in Foliated Field Theory}

\author{Ryan C. Spieler}
\affiliation{Department of Physics and Weinberg Institute for Theoretical Physics, University of Texas at Austin\\
 Austin TX 78712, USA}

\emailAdd{rcspieler@utexas.edu}

\abstract{Fracton phases of matter are gapped phases of matter that, by dint of their sensitivity to UV data, demand non-standard quantum field theories to describe them in the IR.  Two such approaches are foliated quantum theory and exotic field theory.  In this paper, we explicitly construct a map from one to the other and work out several examples.  In particular, we recover the equivalence between the foliated and exotic fractonic BF theories recently demonstrated at the level of operator correspondence.  We also demonstrate the equivalence of toric code layers and the anisotropic model with lineons and planons to the foliated BF theory with one and two foliations, respectively.  Finally, we derive new exotic field theories that provide simple descriptions of hybrid fracton phases from foliated field theries known to do so.  Our results both provide new examples of exotic field theories and pave the way toward their systematic construction from foliated field theories.}

\begin{document}
\maketitle
\flushbottom

\section{Introduction}
\label{sec:intro}

Recent years have seen a sweeping generalization of the concept of global symmetry applied throughout theoretical physics \cite{HighFSym,SnowmassGenSym,GenSymCondMat}.  Approximately in parallel, fracton phases of matter \cite{HermNandRev,PretkoRev} emerged as phenomena requiring a framework.  Fractons are quasiparticles that cannot move.  Depending on the specifics of the model in which they emerge, they might be able to form mobile bound states.  These mobile bound states often only have mobility in some directions - hence names such as lineon and planon.  Fracton phases of matter are gapped phases of matter that possess these excitations \footnote{There are also gapless systems with particles with reduced mobility, see \cite{HermNandRev,PretkoRev,RadGromRev} and references therein.}.  Like topological phases of matter (see, for example, \cite{CMQFT} for a treatment), they have a ground state degeneracy that is robust to local perturbations and reliant on the topology of the manifold on which the system lives (i.e. the presence of non-contractible cycles).  Unlike in topological phases of matter, the ground state degeneracy is often subextensive and depends sensitively on the number of edges in the lattice.  This, combined with the aforementioned mobility constraints, reflects a peculiar sensitivity of the long distance ``universal" physics to short distance data.  These features make it clear that fracton phases cannot be straightforwardly described using topological quantum field theory, unlike their comparatively standard counterparts.  This raises a natural question: what describes the long distance physics of fractons?

Subsystem symmetry \cite{SS1,SS2,SS3} plays a key role in the structure of fracton phases.  On the lattice, a subsystem symmetry only acts on parts of the lattice.  In the continuum, the defects associated with subsystem symmetries cannot undergo arbitrary (homological) deformations \footnote{Any statements made about invariance under deformations supposes that the symmetry defect does not cross an operator charged under the defect.}, but must stay pinned to some submanifold of the space on which the system lives.  Field theories with subsystem symmetry have been thoroughly examined (c.f. \cite{SS1,SS2,SS3}), inspired by the fact that the effective field theory \cite{XCubeQFT,SS3} for a prototypical fracton model called the X-Cube \cite{XCube} possesses subsystem symmetries.  This theory is of a BF type, similar to the theories that describe topological phases \cite{CMQFT}, but the subsystem symmetry is a crucial difference. The restriction on the deformation of symmetry defects is a probe limit of the restricted mobility of the excitations.  Moreover, the mixed 't-Hooft anomalies of the subsystem symmetries force the theories to have a non-trivial Hilbert space.  After suitable regularization, this accounts for the bizarre ground state degeneracy of fracton phases.  These field theories have been dubbed ``exotic" field theories by \cite{Branes}.    

Another line of work showed that some fracton phases, such as the X-Cube, are only sensible on manifolds admitting a foliation \cite{Foliations}.  This lead to a low energy description of fracton phases in terms of foliated quantum field theory \cite{SAW,FoliatedQFT,ElectricMagneticModels}.  The key to foliated quantum field theories is the foliation one form $e$ that is perpendicular to the leaves of the foliation.  By coupling fields to $e$, one can engineer field theories whose defects have similar mobility restrictions.  For example, with three foliations, one can have defects that stay on leaves (planons for flat foliations) or defects that must live at the intersection of two leaves (lineons for flat foliations).  In \cite{SAW} the authors demonstrate that the lattice model for the foliated field theory is in the same phase as the lattice model for the exotic field theory for several gapped systems.  Moreover, \cite{ElectricMagneticModels} remarks that foliated theories with a flat two form gauge field appear to be related to symmetric, off diagonal tensor gauge theories such as those in the exotic field theories.  

Recently, \cite{FoliatedExotic} made the correspondence between foliated and exotic theories explicit for some well known models.  It did so by specifying which operator in the foliated theory corresponded to which operator in the exotic field theory.  This work seeks a constructive approach to the duality outlined therein.  Specifically, we show that by integrating out certain fields in the foliated theory, one can address the constraints imposed in such a way that the exotic theory falls out.  Our procedure is a continuum version of the map in \cite{SAW}.  Before outlining the procedure, let's recapitulate foliated and exotic field theory in some more detail. 
\subsection{Foliated Field Theory}
\label{subsec:introfol}
Foliated quantum field theory was introduced in \cite{SAW} as a continuum description of certain fracton models, called foliated fracton phases.  By foliated fracton phase, we mean a fracton phase that can be defined on manifolds that admit a foliation \cite{Foliations}.  We consider a codimension one foliation of a manifold, that is to say a decomposition along a certain of the manifold into a union of codimension one manifolds called leaves.  The foliation is tracked by the foliation one form $e = e_\mu dx^\mu$.  The vectors $v^\mu$ tangent to each leaf of the foliation must satisfy $v^\mu e_\mu = 0$, which is what we mean when we say the foliation one form is perpendicular to the leaves of the foliation.  The foliation one form must be supported on the entire manifold, and must satisfy $e \wedge de = 0$.  We will assume that $de=0$, though generally $de = e\wedge \omega$, where $\omega$ is called the Godbilon-Vey invariant of the foliation.  Since we suppose $de=0$, we can locally write $e =df$, for some zero form $f$.  Integrating over $e$ counts the number of leaves in the foliation after one introduces a lattice regularization.  Note that the consistency conditions are agnostic to rescaling $e$ i.e. $e \sim \gamma e$, for some zero form $\gamma$.  Thus, we can choose the number of leaves in the foliation, and the IR theory is sensitive to this choice.  This sort of UV/IR mixing is ubiquitous in fractonic field theories.  We can have multiple foliations, leading to multiple foliation one forms $e^A$.  

The foliated field theories we study in this paper are variants of the following BF theory:
\begin{equation}
    \mathcal{L} = \frac{iN}{2\pi} [\sum_A (e^A \wedge B^A \wedge dA^A - e^A \wedge b \wedge A^A) + b\wedge da]
\end{equation}
    Let $d$ be the number of spacetime dimensions.  Here, $a$ is a one form gauge field (albeit with nonstandard gauge transformations, detailed below), $b$ is a $d-2$ form gauge field, $A^A$ are 1 form gauge fields, and $B^A$ are $d-3$ form gauge fields \footnote{By p form gauge field, we of course mean a connection that is locally a p form.  Generally, they are not globally p forms.  The same caveat applies to the fields that parameterize the gauge redundancy.}.  Inspecting the first term of the above Lagrangian informs us that $A^A$ and $B^A$ only have components perpendicular to the direction of $e$, so terms involving foliated gauge fields describe physics on the leaves of a foliation.  The third term is (up to the above subtlety in the gauge transformations) the usual BF theory, and the middle term couples the two.  While we view $A^A$, $B^A$, $a$, and $b$ as dynamical fields, we only consider static foliations in this work.  The gauge holonomies, which are probe limits of the excitations in the system \footnote{To save words, we will often be sloppy about this and refer to the holonomy by the excitation of which it is a probe limit.  For example, we will often call holonomies on manifolds that cannot be deformed into spatial directions fractons.}, are the gauge invariant observables in the theory.  They care about the leaves of the foliation, and can naturally live on the leaves or their intersections (considering leaves of different foliations, of course).  By our mapping onto an exotic field theory, we will see this explicitly throughout the paper.  One can also analyze the gauge holonomies directly in the foliated picture, see \cite{SAW,FoliatedQFT,ElectricMagneticModels,FoliatedExotic}.  To calculate the ground state degeneracy, one can solve Gauss' law as in \cite{FoliatedExotic} and counting the number of modes in the solution.  Doing so involves counting the number of leaves in the foliation, so we obtain a sub-extensive result.    

    Let us pause to clarify our notation, especially as compared to the literature.  What we denote as $A^A \wedge e^A$ is denoted $A^A$ in \cite{FoliatedQFT} and $B^A$ in \cite{ElectricMagneticModels}.  Similarly, what we denote as $B^A \wedge e^A$ is denoted $B^A$ in \cite{FoliatedQFT} and $A^A$ in \cite{ElectricMagneticModels}.  In this regard, our notation is most similar to \cite{SAW,FoliatedExotic}.  Compared to the latter, we will differ by some signs, since we follow the conventions of \cite{SAW} when writing our actions.  The exceptions to this are that we work in Euclidean signature and use $\lbrace A \rbrace$ to index foliations rather than $\lbrace k \rbrace$ \footnote{We have other plans for the letter $k$.}.    
\subsection{Exotic Field Theory}
\label{subsec:introexotic}

Exotic field theories are field theories that make subsystem symmetries apparent by dint writing fields that transform under discrete subgroups of the rotation group.  It is best to illustrate with an example:
\begin{equation}
    \mathcal{L} = \frac{iN}{4\pi}[A_{ij}(\partial_\tau \hat{A}^{ij} - \partial_k \hat{A}_\tau^{k(ij)}) + A_\tau(\partial_i \partial_j \hat{A}^{ij})]
\end{equation}
Here, $i,j,k$ are distinct indices and $i$ and $j$ are symmetric.  Moreover, spatial indices label transformation under $S_4$, the group orientation preserving rotations of a cube by $\pi/2$, rather than transformation under $SO(3)$.  Details on the representation theory of $S_4$ can be found in \cite{SS3}.  As discussed in detail below, this theory's gauge invariant operators are gauge holonomies.  Thanks to its non-standard gauge transformations, they are constrained to live on lines and planes.  Computing the ground state degeneracy amounts the counting the operators, which amounts to counting planes.  This diverges, and must be regularized by placing the theory on a cubic lattice.  This procedure yields the trademark subextensive ground state degeneracy.    

Throughout this paper, fields in exotic field theories are written in uppercase letters.  This will occassionally result in the need to capitalize fields when mapping from foliated to exotic field theory \footnote{Notably, this means we operate well outside the convention in which lowercase fields are dynamical (i.e. they appear in the path integral measure) and uppercase fields are background fields.}.
\subsection{Outline of the Paper}
\label{subsec:introoutline}

In this paper, we exploit a simple fact to obtain exotic field theories from foliated field theories - the exotic field theories are foliated theories with the constraints that relate foliated and standard gauge fields imposed.  We do so by integrating out the time components of the fields in the middle term of the first Lagrangian above.

The remainder of the paper is structured as follows.  We begin with the foliated BF theory in 2+1 dimensions.  For 2 foliations, our procedure yields the exotic BF theory in \cite{SS1} and the correspondence between operators in \cite{FoliatedExotic}.  We then move to the foliated BF theory in 3+1 dimensions.  For one foliation, our procedure yields a field theory for a stack of toric codes (2+1 dimensional $\mathbb{Z}_N$ gauge theories \footnote{In condensed matter literature, and in this paper, one often refers to $\mathbb{Z}_N$ gauge theory as a toric code, in homage to the lattice model introduced in \cite{TC}.}).  We have not seen this field theory in the literature before.  For two foliations, we find the anisotropic theory with lineons, which has appeared in several places \cite{ExiFol,FCC,XCubeBndy}.  For three foliations, we obtain the exotic BF theory of the X-Cube from \cite{XCubeQFT,SS3} and the operator correspondence from \cite{FoliatedExotic}.  We then move to models obtained by coupling the foliated BF theory to additional gauge fields.  Our procedure yields novel theories that couple exotic gauge fields to conventional gauge fields and exhibit key physics of hybrid fracton phases introduced in \cite{Hybrid}, namely that fractonic (reduced mobility) excitations can fuse to mobile excitations and vice-versa.  Our work is an important step in constructing a map from foliated to exotic field theories, which could provide a systematic way to uncover further exotic field theories.

\section{Foliated BF Theory in 2+1 Dimensions}
\label{sec:3d}

In this section, we map the foliated BF theory in 2+1 dimensions to an exotic theory.  We obtain the exotic BF theory first described in \cite{SS1}.  Moreover, in mapping from the foliated theory to the exotic theory, we identify fields in such a way that we naturally rederive the field dictionary discussed in \cite{FoliatedExotic}.  For the sake of keeping the present work self contained, we discuss the exotic BF theory in detail, describing its gauge invariant operators and deriving and interpreting its ground state degeneracy.  Much of this discussion follows \cite{SS1}.   We recover the map from fields in the foliated theory to fields in the exotic theory that \cite{FoliatedExotic} discusses as a consequence of our procedure.

We consider the foliated BF theory in 2+1 dimensions with two foliations $e^1 = dx$ and $e^2 = dy$.  The Lagrangian is
\begin{equation}
    \label{eq:3dfol}
    \mathcal{L} = \frac{iN}{2\pi}[dx\wedge B^x \wedge dA^x + dy\wedge B^y \wedge dA^y - dx\wedge b \wedge A^x - dy\wedge b\wedge A^y + b\wedge da ]
\end{equation}
Let us take roll of the fields involved.  $a$ is a one form gauge field with gauge redundancy 
\begin{equation}
    \label{eq:3dgt1}
    a \sim a + d\lambda_0 - \sum_A \lambda^A dx^A,
\end{equation}
$b$ is a one form gauge field with gauge redundancy
\begin{equation}
    \label{eq:3dgt2}
    b \sim b + d\mu_0,
\end{equation}
$A^A \wedge dx^A$ is a foliated 1+1 form gauge field with redundancy
\begin{equation}
    \label{eq:3dgt3}
    A^A \wedge dx^A \sim A^A \wedge dx^A + d\lambda^A \wedge dx^A,
\end{equation}
and $B^A$ is a 0 form field with redundancy 
\begin{equation}
    \label{eq:3dgt4}
    B^A \sim B^A + 2\pi m^A -\mu_0,
\end{equation}
where $m^A$ is an integer valued function of $x^A$ alone.  The zero form gauge parameters are compact; they are $2\pi$ periodic.  Intuitively, the first two terms describe decoupled layers of 1+1 dimensional $\mathbb{Z}_N$ gauge theories, the third term describes 2+1 dimensional $\mathbb{Z}_N$ gauge theory, and the middle two terms describe the coupling between the two sectors.  That BF theory is a $\mathbb{Z}_N$ gauge theory is made clear in \cite{5brane,BS}.  One can also gleam this from the quantization of various quantities.  Specifically:
\begin{equation}
    \oint_{\mathcal{C}^{(1)}} b \in \frac{2\pi}{N}\mathbb{Z},
\end{equation}
where $\mathcal{C}^{(1)}$ is a closed one-manifold,
\begin{equation}
    \oint_{\mathcal{C}^\tau} a \in \frac{2\pi}{N}\mathbb{Z},
\end{equation}
where $\mathcal{C}^\tau$ is a closed curve around the $\tau$ cycle,
\begin{equation}
    \oint_{\mathcal{S}^A} A^A \wedge dx^A \in \frac{2\pi}{N}\mathbb{Z},
\end{equation}
where $\mathcal{S}^A$ is a strip whose boundary components are on leaves of the foliation defined by $dx^A$, and 
\begin{equation}
    B^x-B^y \in \frac{2\pi}{N}\mathbb{Z}.
\end{equation}
We now turn to the defects and operators in the theory.  The theory has the defect
\begin{equation}
    \label{eq:2fol3dop1}
    W(x,y) = \exp[i\oint_{\mathcal{C}^\tau}a],
\end{equation}
which we interpret as a fracton at $(x,y)$.  There is a local operator:
\begin{equation}
    \label{eq:2fol3dop2}
    W_e(x,y) = \exp[i(B^x(x,y)-B^y(x,y))], 
\end{equation}
and strips:
\begin{multline}
    \label{eq:2fold3dop3}
    W_{m,x}(x_1,x_2) = \exp[i\int_{x_1}^{x_2} \oint A^x\wedge dx + d(a_x dx)];\\ W_{m,y}(y_1,y_2) = \exp[i\int_{y_1}^{y_2} \oint A^y\wedge dy + d(a_y dy)].
\end{multline}
Courtesy of the flux quantization:
\begin{equation}
    W(x,y)^N = W_e(x,y)^N = W_{m,x}(x_1,x_2)^N = W_{m,y}(y_1,y_2)^N = 1.
\end{equation}

We now extract the exotic theory from the foliated field theory.  Expanding the Lagrangian in components gives
\begin{multline}
    \label{3dfolcomp}
    \mathcal{L} = \frac{iN}{2\pi}[\sum_{A=1}^2 (\delta_i^A B^A \partial_\tau A_j^A + b_i \partial_\tau a_j)\epsilon^{ij} + \sum_A A_\tau^A \delta_i^A (-\partial_j B^A + b_j)\epsilon^{ij}\\ - a_\tau \partial_i b_j \epsilon^{ij} + b_\tau (\partial_i a_j - \sum_A \delta_i^A A_j^A)\epsilon^{ij}
\end{multline}
The second and fourth collections of terms are interesting - they relate the standard gauge fields to the foliated gauge fields.  This is exactly the sort of thing that should be encoded in the exotic field theory.  Thus, we integrate out the corresponding Lagrange multipliers $A_\tau^x$, $A_\tau^y$, and $b_\tau$.  Integrating out $A_\tau^A$ simply relates $b$ and $B$.  Integrating out $b_\tau$ gives the equation
\begin{equation}
    \label{eq:3dconstr1}
    \partial_x a_y - \partial_y a_x + A_x^y - A_y^x = 0
\end{equation}
Let's define 
\begin{equation}
    \label{eq:3ddef1}
    A_{ij} = A_j^i + \partial_j a_i
\end{equation}
Note that satisfying the above constraint amounts to demanding that
\begin{equation}
    \label{eq:3dsym}
    A_{xy} = A_{yx},
\end{equation}
the correct behavior for a symmetric tensor gauge field.  Upon solving for $A_j^i$ in terms of $A_{ij}$, solving for $b_i$ in terms of $\partial_i B^k$, relabeling $a_\tau = A_\tau$ and defining
\begin{equation}
    \label{eq:3ddef2}
    \phi^{xy} = B^x - B^y,
\end{equation}
we obtain
\begin{equation}
    \label{eq:3dexotic}
    \mathcal{L} = \frac{iN}{2\pi}\phi^{xy}(\partial_\tau A_{xy}-\partial_x\partial_y A_\tau),
\end{equation}
which is precisely the BF presentation of the exotic $\mathbb{Z}_N$ gauge theory in 2+1 dimensions.  This theory is discussed in detail in \cite{SS1,TimelikeSymmetry,FoliatedExotic}.  For the sake of being self-contained, let us examine its main features.  The Lagrangian has the gauge redundancy
\begin{equation}
    \label{eq:3dgt5}
    A_\tau \sim A_\tau + \partial_\tau \alpha; A_{xy} \sim A_{xy} + \partial_x \partial_y \alpha
\end{equation}
\begin{equation}
    \label{eq:3dgt6}
    \phi^{xy} \sim \phi^{xy} + 2\pi n^{x} + 2\pi n^y,
\end{equation}
where the gauge parameters are related to those in the foliated theory by
\begin{equation}
    \label{eq:3ddict}
    \alpha = \lambda_0 ; n^x = m^x; n^y = -m^y. 
\end{equation}
One can show the following quantities to be quantized:
\begin{equation}
    \oint d\tau A_\tau \in \frac{2\pi}{N}\mathbb{Z}
\end{equation}
\begin{equation}
    \int_{x_1}^{x_2}dx \oint (d\tau \partial_x A_\tau + dy A_{xy}) \in \frac{2\pi}{N} \mathbb{Z}
\end{equation}
\begin{equation}
    \int_{y_1}^{y_2}dy \oint (d\tau \partial_y A_\tau + dx A_{xy}) \in \frac{2\pi}{N} \mathbb{Z}
\end{equation}
\begin{equation}
    \phi^{xy} \in \frac{2\pi}{N}\mathbb{Z}
\end{equation}
Let us discuss the global symmetry.  There is a defect
\begin{equation}
    \label{eq:3dfract}
    W(x,y) = \exp[i\oint d\tau A_\tau(x,y)],
\end{equation}
which we interpret as a fracton located at $(x,y)$.  There are two types of global symmetries.  The $\mathbb{Z}_N$ electric symmetry is generated by
\begin{equation}
    \label{eq:3dvo}
    W_e(x,y) = \exp[i \phi^{xy}(x,y)],
\end{equation}
whereas the $\mathbb{Z}_N$ magnetic symmetry is generated by
\begin{equation}
    \label{eq:3dstrips}
    W_{m,x}(x_1,x_2) = \exp[i\int_{x_1}^{x_2}dx\oint dy A_{xy}]; W_{m,y}(y_1,y_2) = \exp[i\int_{y_1}^{y_2}dx\oint dx A_{xy}].
\end{equation}
As a consequence of the quantized fluxes:
\begin{equation}
    W(x,y)^N = W_e(x,y)^N = W_{m,x}(x_1,x_2)^N = W_{m,y}(y_1,y_2)^N = 1.
\end{equation}
Thus, we see that the operators/defects \cref{eq:3dfract,eq:3dvo,eq:3dstrips} are the exotic counterparts of \cref{eq:2fol3dop1,eq:2fol3dop2,eq:2fold3dop3} in the foliated field theory.  We remark that these exhaust the symmetries if the periodicity of the torus alligns with the foliations.  We only consider such an untwisted torus in this paper.  Thorough discussion of twisting the boundary conditions can be found in \cite{TwistSym}.  These symmetries give the ground state degeneracy, which appears in the field theory as the dimension of the Hilbert space.  It follows from the canonical commutation relations that \footnote{Algebras of this sort signal a mixed 't Hooft anomaly between the two symmetries.}
\begin{equation}
    \label{3danomaly1}
    W_e(x,y)W_{m,x}(x_1,x_2) =\exp[\frac{2\pi i}{N}]W_{m,x}(x_1,x_2)W_e(x,y) ; x_1<x<x_2
\end{equation}
\begin{equation}
    \label{3danomaly2}
    W_e(x,y)W_{m,y}(y_1,y_2) =\exp[\frac{2\pi i}{N}]W_{m,y}(y_1,y_2)W_e(x,y) ; y_1<y<y_2
\end{equation}
As detailed in \cite{SS1}, when regularized on an $L_x\times L_y$ square lattice, we obtain the ground state degeneracy
\begin{equation}
    \label{eq:3dgsd}
    GSD = N^{L_x+L_y-1}.
\end{equation}
Notably, unlike in a topological order, this ground state degeneracy depends on local operators $W_e$.  In fact, it is not generally robust, as \cite{SS1} discusses in detail.  This is consistent with no-go theorems for fracton order in 2+1 dimensions \cite{DegBound,DefectNet}.  

\section{Foliated BF Theory in 3+1 Dimensions}
\label{sec:4d}

In this section, we obtain exotic field theories from the foliated BF theory in 3+1 dimensions with one, two, and three foliations.  For one foliation, we find that the theory is equivalent to the toric code layers.  Our procedure yields a presentation of this theory that we have not seen in literature, and we analyze it in detail.  For two foliations, the theory is equivalent to the anisotropic theory with lineons and planons.  We find the presentation of the theory studied in \cite{ExiFol,FCC,TimelikeSymmetry} and analyze it in detail.  For three foliations, the theory is equivalent to the exotic presentation of the X-Cube model studied in \cite{XCubeQFT,SS3}.  We discuss this theory in detail.  It is worth noting that we recover the map from fields in the foliated theory to fields in the exotic theory that \cite{FoliatedExotic} discusses as a consequence of our procedure.  These equivalences between foliated and exotic theories were shown on the lattice in \cite{SAW}.

Since we will examine multiple choices of foliation(s), we begin by examining the foliated BF theory in 3+1 dimensions generally.  The Lagrangian is
\begin{equation}
    \label{eq:4dfol}
    \mathcal{L} = \frac{iN}{2\pi}[\sum_A (e^A \wedge B^A \wedge dA^A - e^A\wedge b\wedge A^A) + b\wedge da] 
\end{equation}
As before, we take note of the fields present.  $a$ is a one form gauge field with the gauge redundancy 
\begin{equation}
    \label{eq:4dgt1}
    a \sim a + d\lambda_0 - \sum_A \lambda^A e^A,
\end{equation}
$b$ is a two form gauge field with the gauge redundancy
\begin{equation}
    \label{eq:4dgt2}
    b \sim b + d\mu_1,
\end{equation}
$A^A\wedge e^A$ are foliated 1+1 form gauge fields with redundancy 
\begin{equation}
    \label{eq:4dgt3}
    A^A\wedge e^A \sim A^A \wedge e^A + d\lambda^A \wedge e^A,
\end{equation}
and $B^A$ are foliated 1 form gauge fields with redundancy
\begin{equation}
    \label{eq:4dgt4}
    B^A \sim B^A + d\chi^A + \beta^A e^A + \mu_1.
\end{equation}
 As before, the zero form gauge parameters $\lambda_0$ and $\lambda^A$ are compact by dint of being $2\pi$ periodic.  $\beta^A$ is a zero form gauge parameter compactified by the identification $\beta^A \sim \beta^A + 2\pi$.  $\mu_1$ is a one form gauge parameter.  It has its own gauge redundancy, since we can redefine it by an exact one form $d\mu_0$.  We compactify $\mu_0$ by forcing it be $2\pi$ periodic.  Intuitively, the first two terms describe decoupled layerings of 2+1 dimensional $\mathbb{Z}_N$ gauge theories, the third term describes 3+1 dimensional $\mathbb{Z}_N$ gauge theory, and the middle two terms describe the coupling between the two sectors.  One can show the following quantities to be quantized:
\begin{equation}
    \oint_{\mathcal{C}^{(2)}} b \in \frac{2\pi}{N}\mathbb{Z},
\end{equation}
where $\mathcal{C}^{(2)}$ is a closed two-manifold,
\begin{equation}
    \oint_{\mathcal{C}^L} a \in \frac{2\pi}{N}\mathbb{Z},
\end{equation}
where $\mathcal{C}^L$ is a one-cycle on the intersection of leaves of different foliations,
\begin{equation}
    \oint_{\mathcal{S}^A} A^A \wedge e^A \in \frac{2\pi}{N}\mathbb{Z},
\end{equation}
where $\mathcal{S}^A$ is a strip whose boundary components are on different leaves of the foliation defined by $e^A$, and
\begin{equation}
    \oint \sum_A q_A B^A \in \frac{2\pi}{N}\mathbb{Z} ; \sum_A q_A = 0; q_A \in \mathbb{Z},
\end{equation}
where $\mathcal{C}^I$ is a one-cycle in the intersection of leaves in each foliation for which $q_A \neq 0$.
Let us now discuss some of the operators/defects in the theory.  The first is 
\begin{equation}
    \label{eq:4dop1}
    W_m = \exp[i\oint_{\mathcal{C}^L} a].
\end{equation}
  For one foliation in ,$e =dz$, $\mathcal{C}^L$ is in the $\tau-x-y$ space.  For two foliations :$e^1 = dx$ and $e^2 = dy$, $\mathcal{C}^L$ is in the $\tau-z$ plane.  For three foliations: $e^1 =dx$, $e^2=dy$, and $e^3 = dz$, $\mathcal{C}^L$ is a in $\tau$ i.e. we have a fracton.  There are electric holonomies:
\begin{equation}
    \label{eq:4dop2}
    W_e = \exp [i\oint_{\mathcal{C}^I} \sum_A q_A B^A]; \sum_A q_A = 0; q_A \in \mathbb{Z}.
\end{equation}
  For one foliation, this holonomy is the identity.  For two and three foliations, we obtain lineons.  There are also magnetic holonomies (written below for flat foliations $e^A = dx^A$):
\begin{equation}
    W_m^A = \exp[i\oint_{\mathcal{S}^A} (A^A \wedge e^A + d(a_A dx^A))],
\end{equation}
where $\mathcal{S}^A$ is a strip whose boundary is the disjoint union of two leaves of the $A^{th}$ foliation.  Thus, this is a planon.  The quantization conditions imply:
\begin{equation}
    \label{eq:4dop3}
    (W_m)^N = (W_e)^N = (W_m^A)^N = 1.
\end{equation}

We remark that lattice models corresponding to this foliated field theory for the foliation choices we discuss below were constructed in \cite{SAW}.  Those authors then obtained more familiar lattice theories by acting with a finite depth local unitary circuit (we note that this keeps the Hamiltonian in the same phase \cite{FDLU}) and treating local stabilizers as constraints.  The constraints we enforce by integrating out Lagrange multipliers are continuum versions of those constraints.  We also remark that this is the first section in which novel exotic theories are obtained, since \cite{ElectricMagneticModels,FoliatedExotic} only discuss the relationship between foliated and exotic field theories for 3 foliations in 3+1 dimensions.
\subsection{One Foliation: Toric Code Layers}
\label{subsec:4d1fol}

Let $e=dz$.  In components, the Lagrangian is
\begin{multline}
    \label{eq:4d1folcomp}
    \mathcal{L} = \frac{iN}{2\pi}[\delta_i^z B_j^z \partial_\tau A_k^z \epsilon^{ijk} + \frac{1}{2}b_{ij}\partial_\tau a_k \epsilon^{ijk} + A_\tau^z \delta_i^z(-\partial_j B_k^z + \frac{1}{2}b_{jk})\epsilon^{ijk} + \frac{1}{2}a_\tau \partial_i b_{jk} \\+ b_{\tau i}(\partial_j a_k - \delta_j^z A_k^z)\epsilon^{ijk} + B_\tau^z \delta_i^z \partial_j A_k^z \epsilon^{ijk}]
\end{multline}
As usual, we integrate out fields that relate foliated and standard gauge fields, that is we integrate out $A_\tau^z$, $b_{\tau x}$, and $b_{\tau y}$.  Since nothing is treated symmetrically, we simply solve for the usual fields in terms of the foliated fields and plug in, giving 
\begin{multline}
    \label{eq:4d1folcomp'}
    \mathcal{L} = \frac{iN}{2\pi}[A_\tau(\partial_z \partial_x B_y^z - \partial_z \partial_y B_x^z +\partial_x b_{yz} + \partial_y b_{zx}) + A_x(\partial_y \partial_z B_\tau^z + \partial_y b_{\tau x} - \partial_\tau \partial_z B_y^z - \partial_z b_{yz})\\ + A_y(\partial_\tau \partial_z B_x^z - \partial_x \partial_z B_\tau^z - \partial_\tau b_{zx} - \partial_x b_{\tau z})]
\end{multline}
We define
\begin{equation}
    \label{eq:4d1fold1}
    \hat{A}_{iz} = \partial_z B_i^z + b_{iz} 
\end{equation}
for $i \in \lbrace x,y \rbrace$, and
\begin{equation}
    \label{eq:4d1fold2}
    \hat{A}_{\tau z} = \partial_z B_\tau^z + b_{\tau z} 
\end{equation}
so that
\begin{equation}
    \label{eq:4d1folgt1}
    \hat{A}_{iz} \sim \hat{A}_{iz} + \partial_i \partial_z \chi^z + \partial_i \mu_z
\end{equation} 
\begin{equation}
    \label{eq:4d1folgt2}
     \hat{A}_{\tau z} \sim \hat{A}_{\tau z} + \partial_\tau \partial_z \chi^z + \partial_\tau \mu_z.
\end{equation}
We can now rewrite the Lagrangian in the following suggestive form
\begin{equation}
    \label{eq:TCstack}
    \mathcal{L} = \frac{iN}{2\pi}[A_\tau (\epsilon^{ij}\partial_i \hat{A}_{jz}) + \epsilon^{ij}A_i(\partial_j \hat{A}_{\tau z} - \partial_\tau \hat{A}_{jz})]
\end{equation}
Note that, upon dimensional reduction in z, we obtain the usual BF theory in 2+1 dimensions, as expected.  One can show the following periods to be quantized:
\begin{equation}
    \oint (d\tau A_\tau + dx A_x + dy A_y) \in \frac{2\pi}{N}\mathbb{Z}
\end{equation}
\begin{equation}
    \int_{z_1}^{z_2}dz\oint (d\tau \hat{A}_{\tau z} + dx \hat{A}_{xz} + dy \hat{A}_{yz}) \in \frac{2\pi}{N}\mathbb{Z}.
\end{equation}

Let us examine the global symmetries of the theory.  It has a defect
\begin{equation}
    \label{eq:TCstackfrac}
    W_m = \exp[i\oint d\tau A_\tau]
\end{equation}
and a magnetic symmetry, generated by the operators
\begin{equation}
    \label{eq:TCstackline}
    W_m^x = \exp[i\oint dx A_x]; W_m^y = \exp[i\oint dy A_y].
\end{equation}
It also has an electric symmetry, generated by
\begin{equation}
    \label{eq:TCstackstrip}
    W_e^x(z_1,z_2) = \exp[i\int_{z_1}^{z_2}dz\oint dx \hat{A}_{xz}]; 
    W_e^y(z_1,z_2) = \exp[i\int_{z_1}^{z_2}dz\oint dy \hat{A}_{yz}]
\end{equation}
We note that the above holonomies are in fact gauge invariant.  $\mu_z$ has the gauge symmetry $\mu_z \sim \mu_z + \partial_z \beta$, where $\beta$ is a zero form.  Thus, the argument of the exponent reduces to $2\pi$ times the difference between winding numbers at different values of z, which give a trivial holonomy.  There is an electric defect that can be defined as expected:
\begin{equation}
    \label{eq:TCelectricdefect}
    W_e(z_1,z_2) = \exp[i\int_{z_1}^{z_2}dz\oint dt \hat{A}_{tz}].
\end{equation}
It follows from the quantized periods that
\begin{equation}
    W_m^N = W_e(z_1,z_2)^N = (W_m^x)^N = (W_m^y)^N = W_e^x(z_1,z_2)^N = W_e^y(z_1,z_2)^N = 1.
\end{equation}
This is consistent with the results from the foliated side of the duality.  \footnote{The electric operators correspond to operators and defects that do not show up in the foliated Lagrangian, so one cannot quantize the fluxes on the foliated side of the duality in the manner reviewed in the appendix.  The quantization on the exotic side can be demonstrated in that manner, and we expect them to match.  We leave a detailed demonstration to further work.  Similar remarks apply to operators constructed from $\hat{A}$ in the anisotropic model with lineons and planons and the hybrid versions of these theories.}

We use these symmetries to determine the ground state degeneracy of the system on $T^3$, represented in the field theory by the dimension of the Hilbert space.  It follows from the canonical commutation relations that the symmetry operators satisfy:
\begin{equation}
    \label{eq:TCstackanomaly}
    W_e^x(z_1,z_2) W_m^y(z) = \exp[\frac{2\pi i}{N}]W_m^y(z)W_e^x(z_1,z_2); z_1<z<z_2
\end{equation}
and similarly for $x \leftrightarrow y$.  This implies
\begin{equation}
    \label{eq:TCstackgsd}
    GSD = N^{2L_z}
\end{equation}
if we discretize the z axis.  This is the result we would expect from dimensional reduction.

We note that fracton orders based on layered toric code appear in \cite{ICS,ICS2,TwistedFoliated}, which analyze them in terms of infinite component K-matrix Chern-Simons theory.  
\subsection{Two Foliations: Anisotropic Theory with Lineons and Planons}
\label{subsec:4d2fol}

Let $e^1 = dx$ and $e^2 = dy$.  In component form, the Lagrangian is
\begin{multline}
    \label{eq:4d2folcomp}
    \mathcal{L} = \frac{iN}{2\pi}[(\delta_i^x B_j^x \partial_\tau A_k^x + \delta_i^y B_j^y \partial_\tau A_k^y + \frac{1}{2}b_{ij}\partial_\tau a_k)\epsilon^{ijk} \\ +A_\tau^x \delta_i^x (-\partial_j B_k^x + \frac{1}{2}b_{jk})\epsilon^{ijk} + A_\tau^y \delta_i^y (-\partial_j B_k^y + \frac{1}{2}b_{jk})\epsilon^{ijk} + \frac{1}{2}a_\tau \partial_i b_{jk}\epsilon^{ijk} \\ b_{\tau i}(\partial_j a_k - \delta_j^x A_k^x - \delta_j^y A_k^y)\epsilon^{ijk} - B_\tau^x \delta_i^x \partial_j A_k^x \epsilon^{ijk} - B_\tau^y \delta_i^y \partial_j A_k^y \epsilon^{ijk}]
\end{multline}
We integrate out $b_{\tau x}$, $b_{\tau y}$, $A_\tau^x$, and $A_\tau^y$.  Let's examine the consequences.  Integrating out $A_\tau^x$ imposes
\begin{equation}
    \label{eq:4d2folconstr1}
    -\partial_y B_z^x + \partial_z B_y^z +b_{yz} = 0
\end{equation}
whereas integrating out $A_\tau^y$ imposes
\begin{equation}
    \label{eq:4d2folconstr2}
    -\partial_z B_x^y + \partial_x B_z^y +b_{zx} = 0
\end{equation}
We do not do any more with these constraints for the time being.  Integrating out $b_{\tau z}$ gives
\begin{equation}
    \label{eq:4d2folconstr3}
    \partial_x a_y - \partial_y a_x - A_y^x + A_x^y = 0.
\end{equation}
We note this is solved by
\begin{equation}
    \label{eq:4d2foldef1}
    A_{xy} = A_y^x + \partial_y a_x
\end{equation}
provided we demand $A_{xy} = A_{yx}$.  This field has the same gauge redundancy as in \eqref{eq:3dgt5}.
Integrating out $b_{\tau x}$ gives
\begin{equation}
    \label{eq:4d2folconstr4}
    \partial_y a_z - \partial_z a_y - A_z^y = 0
\end{equation}
and integrating out $b_{\tau y}$ gives
\begin{equation}
    \label{eq:4d2folconstr5}
    \partial_z a_x - \partial_x a_z + A_z^x = 0.
\end{equation}
Upon solving for $A_y^x$, $A_x^y$, $A_z^x$, $A_z^y$, $b_{zx}$, and $b_{yz}$ using the constraints, rewriting $a_z=A_z$ and defining the following:
\begin{equation}
    \label{eq:4d2folsdef2}
    \hat{A}_\tau^{xy} =  B_\tau^x - B_\tau^y
\end{equation}
\begin{equation}
    \label{eq:4d2foldef3}
    \hat{A} = \partial_x B_y^x - \partial_y B_x^y - b_{xy}
\end{equation}
\begin{equation}
    \label{eq:4d2foldef4}
    \hat{A}_z^{xy} = B_z^x - B_z^y
\end{equation}
so that
\begin{equation}
    \label{eq:4d2folgt1}
    \hat{A}_\tau^{xy} \sim \hat{A}_\tau^{xy} + \partial_\tau \hat{\alpha}^{xy}
\end{equation}
\begin{equation}
    \label{eq:4d2folgt2}
    \hat{A} \sim \hat{A} + \partial_x \partial_y \hat{\alpha}^{xy}
\end{equation}
\begin{equation}
    \label{eq:4d2folgt3}
    \hat{A}_z^{xy} \sim \hat{A}_z^{xy} + \partial_z \hat{\alpha}^{xy},
\end{equation}
where the gauge parameters in the exotic field theory are related to those in the foliated theory as
\begin{equation}
    \label{eq:4d2foldict}
    \alpha = \lambda_0 ; \hat{\alpha}^{xy} = \chi^x - \chi^y,
\end{equation}
we obtain
\begin{multline}
    \label{eq:AnisExotic}
    \mathcal{L} = \frac{iN}{2\pi}[A_{xy}(\partial_\tau \hat{A}_z^{xy} - \partial_z \hat{A}_\tau^{xy}) + A_z (\partial_\tau \hat{A} - \partial_x \partial_y \hat{A}_\tau^{xy}) \\ + A_\tau ( \partial_x \partial_y \hat{A}_z^{xy} -\partial_z \hat{A})].
\end{multline}
Let us analyze this theory.  The following periods are quantized:
\begin{equation}
    \oint (d\tau A_\tau + dz A_z) \in \frac{2\pi}{N}\mathbb{Z}
\end{equation}
\begin{equation}
    \oint (d\tau \hat{A}_\tau^{xy} + dz \hat{A}_z^{xy}) \in \frac{2\pi}{N}\mathbb{Z} 
\end{equation}
\begin{equation}
    \int_{x_1}^{x_2}dx\oint(d\tau \partial_x A_\tau + dy A_{xy}) \in \frac{2\pi}{N}\mathbb{Z}
\end{equation}
\begin{equation}
    \int_{y_1}^{y_2}dy\oint(d\tau \partial_y A_\tau + dx A_{xy}) \in \frac{2\pi}{N}\mathbb{Z}
\end{equation}
\begin{equation}
    \int_{x_1}^{x_2}dx\oint(d\tau \partial_x \hat{A}_\tau^{xy} + dy \hat{A}) \in \frac{2\pi}{N}\mathbb{Z}
\end{equation}
\begin{equation}
    \int_{y_1}^{y_2}dy\oint(d\tau \partial_y \hat{A}_\tau^{xy} + dx \hat{A}) \in \frac{2\pi}{N}\mathbb{Z}
\end{equation}

The theory has the defects:
\begin{equation}
    \label{eq:Anisfrac}
    W_m = \exp[i\oint d\tau A_\tau]; W_e = \exp[i\oint d\tau \hat{A}_\tau]
\end{equation}
It has a magnetic symmetry, generated by
\begin{equation}
    \label{eq:Anislinem}
    W_m(x,y) = \exp[i\oint dz A_z],
\end{equation}
which is a lineon in the z direction and
\begin{equation}
    \label{eq:Anisstripm}
    W_m(x_1,x_2) = \exp[i\int_{x_1}^{x_2}\oint dy A_{xy}]; W_m(y_1,y_2) = \exp[i\int_{y_1}^{y_2}dy\oint dx A_{xy}],
\end{equation}
providing planons in the y-z and x-z planes, respectively \footnote{Note that the equation $\partial_z A_{xy} - \partial_x \partial_y A_z = 0$ allows us to deform these operators in the stated planes.  We write the special case \eqref{eq:Anisstripm} in the paper since it is what contributes to the ground state degeneracy.}.  It also has an electric symmetry, generated by
\begin{equation}
    \label{eq:Anislinee}
    W_e(x,y) = \exp[i\oint dz \hat{A}_z^{xy}],
\end{equation}
giving a lineon in the z direction and 
\begin{equation}
    \label{eq:Anisstripe}
    W_e(x_1,x_2) = \exp[i\int_{x_1}^{x_2}dx\oint dy \hat{A}]; W_e(y_1,y_2) = \exp[i\int_{y_1}^{y_2}dy\oint dx \hat{A}], 
\end{equation}
providing planons in the y-z and x-z planes, respectively \footnote{Note that the equation $\partial_z \hat{A} - \partial_x \partial_y \hat{A}_z^{xy} = 0$ allows us to deform these operators in the stated planes.  We write the special case \eqref{eq:Anisstripe} in the paper since it is what contributes to the ground state degeneracy.}.  The quantized periods imply:
\begin{multline}
    W_m^N = W_e^N = W_m(x,y)^N =W_e(x,y)^N = W_m(x_1,x_2)^N = W_m(y_1,y_2)^N \\ = W_e(x_1,x_2)^N =W_m(y_1,y_2)^N=1,
\end{multline}
consistent with results on the foliated side of the duality.

One can obtain the ground state degeneracy on $T^3$ from the symmetries of the theory.  The canonical commutation relations imply: 
\begin{equation}
    \label{eq:Anisanomaly1}
    W_e(x,y) W_m(x_1,x_2) = \exp[\frac{2\pi i}{N}] W_m(x_1,x_2)W_e(x,y); x_1<x<x_2
\end{equation}
\begin{equation}
    \label{eq:Anisanomaly2}
     W_e(x,y) W_m(y_1,y_2) = \exp[\frac{2\pi i}{N}] W_m(y_1,y_2)W_e(x,y); y_1<y<y_2
\end{equation}
When accounting for constraints (detailed in \cite{TimelikeSymmetry}) and regularizing the x-y plane on on an $L_x \times L_y$ square lattice gives $L_x+L_y-1$ operators.  Note that we could have done this calculation exchanging $e$ and $m$, resulting in another $L_x+L_y-1$ operators and giving a ground state degeneracy of
\begin{equation}
    \label{eq:Anisgsd}
    GSD = N^{2L_x+2L_y-2}.
\end{equation}

\subsection{Three Foliations: The X-Cube Model}
\label{subsec:4d3fol}

Let $e^1 = dx$, $e^2 = dy$, and $e^3=dz$.  In components, the Lagrangian is
\begin{multline}
    \label{eq:4d3folcomp}
    \mathcal{L} = \frac{iN}{2\pi}[(\sum_A \delta^A_i B_j^A \partial_\tau A^A_k + \frac{1}{2}b_{ij}\partial_\tau a_k)\epsilon^{ijk} + \sum_A A_\tau^A \delta_i^A (-\partial_j B_k^A + \frac{1}{2}b_{jk})\epsilon^{ijk} \\ + \frac{1}{2}a_\tau \partial_i b_{jk}\epsilon^{ijk} + b_{\tau i} (\partial_j a_k - \sum_A \delta^A_j A_k^A)\epsilon^{ijk} - \sum_A B_\tau^A \delta_i^A \partial_j A_k^A \epsilon^{ijk}],
\end{multline}
so that integrating out $A_\tau^A$ imposes, for all $A$
\begin{equation}
    \label{eq:4d3folconstr1}
    \delta_i^A (-\partial_j B_k^A + \frac{1}{2}b_{jk})\epsilon^{ijk} = 0.
\end{equation}
As before, we use this to solve for $b$ and plug directly into the Lagrangian. Integrating out $b_{\tau a}$ gives
\begin{equation}
    \label{eq:4d3folconstr2}
    (\partial_j a_k - \sum_A \delta^A_j A_k^A)\epsilon^{ijk} = 0.
\end{equation}
We introduce the symmetric gauge field as before, defining
\begin{equation}
    \label{eq:4d3foldef1}
    A_{ij} = A_j^i + \partial_j a_i
\end{equation}
and demanding that $A_{ij}=A_{ji}$.  We further define
\begin{equation}
    \label{eq:4d3foldef2}
    \hat{A}_\tau^{k(ij)} = B_\tau^i - B_\tau^j
\end{equation}
\begin{equation}
    \label{eq:4d3foldef3}
    \hat{A}^{ij} = B_k^i - B_k^j.
\end{equation}
These fields have the gauge redundancy 
\begin{equation}
    \label{eq:4d3folgt1}
    A_{ij} \sim A_{ij} + \partial_i \partial_j \alpha
\end{equation}
\begin{equation}
    \label{eq:4d3folgt2}
    \hat{A}^{ij} \sim \hat{A}^{ij} + \partial_k \hat{\alpha}^{k(ij)}
\end{equation}
\begin{equation}
    \label{eq:4d3folgt3}
    \hat{A}_\tau^{k(ij)} \sim \hat{A}_\tau^{k(ij)} + \partial_\tau \hat{\alpha}^{k(ij)},
\end{equation}
where the gauge parameters in the exotic field theory can be written in terms of those in the foliated theory as
\begin{equation}
    \label{eq:4d3foldict}
    \alpha = \lambda_0 ; \hat{\alpha}^{k(ij)} = \chi^i - \chi^j.
\end{equation}
Writing $a_\tau = A_\tau$, solving for $b_{ij}$ and $A_i^A$ as indicated above, and plugging into the remaining Lagrangian gives
\begin{equation}
    \label{eq:Xcube}
    \mathcal{L} = \frac{iN}{4\pi}[A_{ij}(\partial_\tau \hat{A}^{ij} - \partial_k \hat{A}_\tau^k) + A_\tau(\partial_i \partial_j \hat{A}^{ij})]
\end{equation}
Note that $i$ and $j$ are symmetric and all values of the indices must be distinct.  The following quantities are quantized:
\begin{equation}
    \oint d\tau A_\tau \in \frac{2\pi}{N}\mathbb{Z}
\end{equation}
\begin{equation}
    \int_{x_1}^{x_2}dx \oint (d\tau \partial_x A_\tau + dy A_{xy} + dz A_{zx}) \in \frac{2\pi}{N}\mathbb{Z}
\end{equation}
\begin{equation}
    \int_{y_1}^{y_2}dy \oint (d\tau \partial_y A_\tau + dx A_{xy} + dz A_{yz}) \in \frac{2\pi}{N}\mathbb{Z}
\end{equation}
\begin{equation}
    \int_{z_1}^{z_2}dz\oint (d\tau \partial_z A_\tau + dx A_{zx} + dy A_{yz}) \in \frac{2\pi}{N}\mathbb{Z}
\end{equation}
\begin{equation}
    \oint (d\tau \hat{A}_\tau^{k(ij)} + dx^k \hat{A}^{ij})\in \frac{2\pi}{N}\mathbb{Z}
\end{equation}
Let us discuss this theory in some detail.  It has a magnetic defect:
\begin{equation}
    \label{eq:Xcubefracm}
    W_m(x,y,z) = \exp[i\oint d\tau A_\tau],
\end{equation}
which cannot be deformed into spatial directions.  We interpret it as a fracton.  The theory has a magnetic dipole symmetry in the yz plane generated by
\begin{equation}
    \label{eq:Xcubeplanon}
    W_m(x_1,x_2,\mathcal{C}^{yz}) = \exp[i\int_{x_1}^{x_2}dx \oint (dy A_{xy} + dz A_{zx})],
\end{equation}
  where $\mathcal{C}^{yz}$ is a 1-cycle in the $yz$ plane.  These operators can be deformed in a gauge invariant way in the $yz$ plane, so we identify them with planons.  The theory also has electric defects:
\begin{equation}
    \label{eq:Xcubefrace}
    \hat{W}_e^k(x,y,z) = \exp[i\oint \hat{A}_\tau^{k(ij)}]
\end{equation}
.  Finally, the theory has an electric tensor symmetry in the x direction generated by
\begin{equation}
    \label{eq:Xcubeline}
    \hat{W}_e^x(y,z) = \exp[i\oint dx \hat{A}^{yz}],
\end{equation}
which gives a lineon in the x direction, since we cannot deform it into other directions in a gauge invariant way.  We can construct similar operators out of $\hat{A}^{xy}$ and $\hat{A}^{zx}$, yielding lineons in the y and z directions, respectively.  It follows from the quantized periods that
\begin{multline}
    W_m(x,y,z)^N = W_m(x_1,x_2,\mathcal{C}^{yz})^N = W_m(y_1,y_2,\mathcal{C}^{zx})^N = W_m(z_1,z_2,\mathcal{C}^{xy})^N \\= \hat{W}_e^x(y,z)^N = \hat{W}_e^y(x,z)^N = \hat{W}_e^z(x,y)^N = 1.
\end{multline}
Thus, we see that \cref{eq:Xcubefracm,eq:Xcubeplanon,eq:Xcubefrace,eq:Xcubeline} are the exotic counterparts to \cref{eq:4dop1,eq:4dop2,eq:4dop3} in the foliated field theory. 

Let us now discuss the calculation of ground state degeneracy on $T^3$ from global symmetries.  The canonical commutation relations imply
\begin{equation}
    \label{eq:Xcubeanomaly1}
    \hat{W}_e^z(x,y) W_m(x_1,x_2,\mathcal{C}_y^{yz}) = \exp[\frac{2\pi i}{N}] W_m(x_1,x_2,\mathcal{C}_y^{yz})\hat{W}_e^z(x,y); x_1<x<x_2
\end{equation}
\begin{equation}
    \label{eq:Xcubeanomaly2}
    \hat{W}_e^z(x,y) W_m(y_1,y_2,\mathcal{C}_x^{xz}) = \exp[\frac{2\pi i}{N}] W_m(y_1,y_2,\mathcal{C}_x^{xz})\hat{W}_e^z(x,y); y_1<y<y_2,
\end{equation}
where $\mathcal{C}_i^{ij}$ is a 1-cycle in the $ij$ plane that winds once in $i$.  When noting a constraint in each sector (discussed in \cite{SS3}), this gives $L_x+L_y-1$ N dimensional spaces from the operators in the x-y plane.  Working similarly in the other planes gives
\begin{equation}
    \label{eq:Xcubegsd}
    GSD = N^{2L_x+2L_y+2L_z-3}
\end{equation}
when the theory is regularized on an $L_x\times L_y \times L_z$ cubic lattice with periodic boundary conditions in all three directions.

\section{The $\mathbb{Z}_{N^2}$ Magnetic Model}
\label{sec:hyb}

The theories we have examined so far all relate the two form $b$ to the foliated one form $B^k$.  For this reason, they are called magnetic models.  One can extend $\mathbb{Z}_N$ by other groups and couple to a $G/\mathbb{Z}_N$ gauge field to produce more general magnetic models \cite{ElectricMagneticModels}.  We label these theories by the group $G$.  One example is the $G=\mathbb{Z}_{N^2}$ magnetic model,which we examine in this section.  We apply the same procedure as before - integrate out the Lagrange multipliers that impose constraints that relate foliated gauge fields to standard gauge fields - to the theory with one, two, and three flat foliations.  For one foliation, we find an exotic field theory for the hybrid toric tode layers.  For two foliations, we find an exotic field theory for a hybrid version of the anisotropic theory with lineons and planons.  For three foliations, we find an exotic version of the fractonic hybrid X-Cube model.  All of the systems were introduced on the lattice in \cite{Hybrid}.  Their characteristic behavior is that they contain both mobile and fractonic excitations, and these two types of excitations fuse into each other.  The relationship between the foliated theory and these systems is documented in \cite{ElectricMagneticModels}.  We have not seen the exotic field theories we uncover in the literature before.  Aside from being an exhibition of our map from foliated to exotic field theories, they are interesting because they provide a simple and generalizable way to capture hybrid fracton physics in exotic field theory.  We detail this, in addition to other features of these theories, on a case by case basis.

Since we will make various choices of foliation(s) throughout this section, give general details about the theory here.  Its Lagrangian is
\begin{equation}
    \label{eq:hybfol}
    \mathcal{L} = \frac{iN}{2\pi}[\sum_A (e^A \wedge B^A \wedge dA^A - e^A\wedge b\wedge A^A) + b\wedge da + b'\wedge da'] - \frac{i}{2\pi}b\wedge da' 
\end{equation}
Let's note the gauge fields present.  $a$ is a one form with gauge field with the redundancy
\begin{equation}
    \label{eq:hybfolgt1}
    a \sim a + d\lambda_0 - \sum_A  \lambda^A e^A,
\end{equation}
b is a two form gauge field with the redundancy 
\begin{equation}
    \label{eq:hybfolgt2}
    b \sim b + d\mu_1,
\end{equation}
$a'$ is a one form gauge field with the gauge redundancy 
\begin{equation}
    \label{eq:hybfolgt3}
    a' \sim a' + d\lambda_0',
\end{equation}
$b'$ is a two form gauge field with the redundancy
\begin{equation}
    \label{eq:hybfolgt4}
    b' \sim b' + d \mu_1',
\end{equation}
$A^A\wedge e^A$ are foliated 1+1 form gauge fields with the redundancy
\begin{equation}
    \label{eq:hybfolgt5}
    A^A\wedge e^A \sim A^A \wedge e^A + d\lambda^A \wedge e^A,
\end{equation}
and $B^A$ are foliated gauge fields with the redundancy
\begin{equation}
    \label{eq:hybfolgt6}
    B^A \sim B^A + d\chi^A + \beta^Ae^A + N\lambda'_1.
\end{equation}
As usual, the zero form gauge parameters $\lambda_0$, $\lambda'_0$, $\lambda^A$,$\beta^A$, and $\chi^A$ are compactified by making them $2\pi$ periodic.  The one form parameters $\mu_1$, $\mu'_1$, and $\lambda'_1$ can all be shifted by exact one forms $d\mu_0$, $d\mu'_0$, and $d\Lambda_0$, respectively.  All of the zero forms so defined are similarly compactified.  One can show the following periods are quantized:
\begin{equation}
    \oint_{\mathcal{C}^{(2)}} b \in \frac{2\pi}{N}\mathbb{Z},
\end{equation}
\begin{equation}
    \oint_{\mathcal{S}^A} A^A \wedge e^A \in \frac{2\pi}{N}\mathbb{Z},
\end{equation}
\begin{equation}
    \oint_{\mathcal{C}^{(1)}} a' \in \frac{2\pi}{N}\mathbb{Z},
\end{equation}
\begin{equation}
    \oint_{\mathcal{C}^I} \sum_A q_A B^A \in \frac{2\pi}{N}\mathbb{Z}; \sum_A q_A = 0,
\end{equation}
\begin{equation}
    \oint_{\mathcal{C}^L} a \in \frac{2\pi}{N^2}\mathbb{Z},
\end{equation}
and
\begin{equation}
    \oint_{\mathcal{C}^{(2)}} b' \in \frac{2\pi}{N^2}\mathbb{Z}.
\end{equation}
All of the manifolds over which these are integrated are as defined above.
The quantities \cref{eq:4dop1,eq:4dop2,eq:4dop3} are still gauge invariant.  However, the above quantization conditions imply that:
\begin{equation}
    (W_m)^{N^2} = (W_e)^N = (W_m^A)^{N^2} = 1.
\end{equation}
Moreover, we also have the line
\begin{equation}
    W_e^{(1)} = \exp[i\oint_{\mathcal{C}^{(1)}}a']
\end{equation}
and the surface
\begin{equation}
    W_m^{(2)} = \exp[i\oint_{\mathcal{C}^{(2)}}b' ].
\end{equation}
Courtesy of the quantization conditions, these satisfy
\begin{equation}
    (W_e^{(1)})^N = (W_m^{(2)})^{N^2} = 1.
\end{equation}
\subsection{One Foliation: Hybrid Toric Code Layers}
\label{subsec:hyb1fol}

Let $e = dz$.  In components, the Lagrangian is
\begin{multline}
    \label{eq:hyb1folcomp}
    \mathcal{L} = \frac{i}{2\pi}[N\delta_i^z B_j^z \partial_\tau A_k^z \epsilon^{ijk} + N\frac{1}{2}b_{ij}\partial_\tau a_k \epsilon^{ijk} + N\frac{1}{2}b'_{\mu\nu}\partial_\rho a'_\sigma \epsilon^{\mu\nu\rho\sigma} - \frac{1}{2}b_{ij}\partial_\tau a'_k \\+ NA_\tau^z \delta_i^z(-\partial_j B_k^z + \frac{1}{2}b_{jk})\epsilon^{ijk} + N\frac{1}{2}a_\tau \partial_i b_{jk} - \frac{1}{2}a'_\tau \partial_i b_{jk} \\+ b_{\tau i}(N\partial_j a_k - N\delta_j^z A_k^z - \partial_j a'_k)\epsilon^{ijk} + NB_\tau^z \delta_i^z \partial_j A_k^z \epsilon^{ijk}]
\end{multline}
As in section \ref{subsec:4d1fol}, we integrate out $A_\tau^z$,$b_{\tau x}$, and $b_{\tau y}$ and define 
\begin{equation}
    \label{eq:hyb1foldef1}
    \hat{A}_{iz} = \partial_z B_i^z + b_{iz} 
\end{equation}
for $i \in \lbrace x,y \rbrace$, and
\begin{equation}
    \label{eq:hyb1foldef2}
    \hat{A}_{\tau z} = \partial_z B_\tau^z + b_{\tau z} ,
\end{equation}
so that these fields have the redundancy in \cref{eq:4d1folgt1,eq:4d1folgt2}.
Solving for $A_k^z$ and $b_{xy}$ using the constraints and plugging the results into the leftover Lagrangian yields
\begin{multline}
    \label{eq:hybtclayer}
    \mathcal{L} = \frac{iN}{2\pi}[A_\tau (\epsilon^{ij}\partial_i A_{jz}) + \epsilon^{ij}A_i(\partial_j \hat{A}_{\tau z} - \partial_\tau \hat{A}_{jz}) + \frac{1}{2}B'_{\mu\nu} \partial_\rho A'_\sigma \epsilon^{\mu\nu\rho\sigma}]\\ -\frac{i}{2\pi}[A'_\tau(\epsilon^{ij} \partial_i A_{jz}) +\epsilon^{ij}A'_i (\partial_j \hat{A}_{\tau z} -\partial_\tau \hat{A}_{jz})]
\end{multline}
The following periods are quantized:
\begin{equation}
    \oint (d\tau A_\tau + dxA_x + dyA_y) \in \frac{2\pi}{N^2}\mathbb{Z}
\end{equation}
\begin{equation}
    \int_{z_1}^{z_2}dz\oint (d\tau \hat{A}_{\tau z} + dx \hat{A}_{xz} + dy \hat{A}_{yz}) \in \frac{2\pi}{N}\mathbb{Z}
\end{equation}
\begin{equation}
    \oint_{\mathcal{C}^{(1)}} A' \in \frac{2\pi}{N}\mathbb{Z}
\end{equation}
\begin{equation}
    \oint_{\mathcal{C}^{(2)}} B' \in \frac{2\pi}{N^2}\mathbb{Z} 
\end{equation}
This theory has more or less the same operators discussed in section \ref{subsec:4d1fol}, with the qualifier stemming from the fact that the quantized periods imply that
\begin{equation}
    W_m^{N^2} = W_e(z_1,z_2)^N = (W_m^x)^{N^2} = (W_m^y)^{N^2} = W_e^x(z_1,z_2)^N = W_e^y(z_1,z_2)^N = 1.
\end{equation}
Moreover, it has a two form symmetry, under which the charged operators are
\begin{equation}
    \label{eq:2form}
    W_m^{(2)}(\mathcal{C}^{(2)}) = \exp[i\oint_{\mathcal{C}^{(2)}}B']
\end{equation}
where $\mathcal{C}^{(2)}$ is a two cycle.  It also has a one form symmetry, under which the charged operators are
\begin{equation}
    \label{eq:1form}
    W_e^{(1)}(\mathcal{C}^{(1)}) = \exp[i\oint_{\mathcal{C}^{(1)}}A'],
\end{equation}
where $\mathcal{C}^{(1)}$ is a one cycle.  Courtesy of the quantization conditions, these obey
\begin{equation}
    (W_m^{(2)})^{N^2} = (W_e^{(1)})^N = 1,
\end{equation}
just as in the foliated field theory.  Moreover, the canonical commutation relations imply 
\begin{equation}
    \label{eq:12anomaly}
    W_e^{(1)}(\mathcal{C}^{(1)}) W_m^{(2)}(\mathcal{C}^{(2)}) = \exp[\frac{2\pi i}{N}I(\mathcal{C}^{(1)},\mathcal{C}^{(2)})]W_m^{(2)}(\mathcal{C}^{(2)})W_e^{(1)}(\mathcal{C}^{(1)})
\end{equation}
where $I(\mathcal{C}^{(1)},\mathcal{C}^{(2)})$ is the intersection number between the two arguments.  This contributes to the ground state degeneracy on $T^3$.  Noting that $\mathcal{C}^{(1)}$ can be generated by a cycle winding x, a cycle winding y, and a cycle winding z.  Moreover, $\mathcal{C}^{(2)}$ can be generated by a cycle wrapping x-y, a cycle wrapping y-z, and a cycle wrapping x-z.  This provides three more N dimensional spaces, so that we have
\begin{equation}
    \label{eq:hybridtcgsd}
    GSD = N^{2L_z + 3}.
\end{equation}  

The terms with prefactor $-\frac{i}{2\pi}$ do not contribute to the ground state degeneracy.  What they \textit{do} contribute are the fusion rules that map N electric planons to a mobile particle and N loops to a magnetic planon. 
 As an example, we demonstrate the former.  The $\hat{A}_{z\tau}$ equation of motion is
\begin{equation}
    \label{eq:tchyb1}
    \epsilon^{ij}\partial_j (NA_i - A'_i) = 0
\end{equation}
Now, imagine fusing $N$ of the line operators:
\begin{equation}
    \label{eq:tchyb2}
    W_e^N = \exp[iN\oint_{\mathcal{C}^{xy}} A] = \exp[iN\int_{\mathcal{D}^{xy}} \Tilde{d}A] = \exp[i\int_{\mathcal{D}^{(2)}} \Tilde{d}A'] = \exp[i\oint_{\mathcal{C}^{(1)}} A'].
\end{equation}
Here, $\Tilde{d}$ is the exterior derivative in the spatial directions, $\mathcal{C}^{xy}$ is a closed curve in the xy plane, $\mathcal{D}^{xy}$ is a region diffeomorphic to a disk in the xy plane such that $\partial \mathcal{D}^{xy} = \mathcal{C}^{xy}$, and $\mathcal{D}^{(2)}$ is a region diffeomorphic to a disk such that $\partial \mathcal{D}^{(2)} = \mathcal{C}^{(1)}$.  The second equality and fourth equalities are from Stokes' theorem and the third equality follows from the equation of motion.  The upshot is that fusing $N$ electric planons gives a mobile particle \footnote{Strictly speaking, this argument, and others like in this paper, is only valid for operators on contactable cycles.  We leave a convincing argument for noncontractable cycles to future work.}.  

This field theory, which we have not seen in this form in the literature, encapsulates the essential physics of the hybrid toric code layers introduced in \cite{Hybrid}.
\subsection{Two Foliations: Hybrid Anisotropic Model with Lineons and Planons}
\label{subsec:hyb2fol}

Let $e^1 =dx$ and $e^2 = dy$.  The Lagrangian is
\begin{multline}
    \label{eq:hyb2folcomp}
    \mathcal{L} = \frac{i}{2\pi}[N\delta_i^x B_j^x \partial_\tau A_k^x \epsilon^{ijk} + N\delta_i^y B_j^y \partial_\tau A_k^y \epsilon^{ijk} + N\frac{1}{2}b_{ij}\partial_\tau a_k \epsilon^{ijk} + N\frac{1}{2}b'_{\mu\nu}\partial_\rho a'_\sigma \epsilon^{\mu\nu\rho\sigma} - \frac{1}{2}b_{ij}\partial_\tau a'_k \\+ NA_\tau^x \delta_i^x(-\partial_j B_k^x + \frac{1}{2}b_{jk})\epsilon^{ijk} + NA_\tau^y \delta_i^y(-\partial_j B_k^y + \frac{1}{2}b_{jk})\epsilon^{ijk} + N\frac{1}{2}a_\tau \partial_i b_{jk} - \frac{1}{2}a'_\tau \partial_i b_{jk} \\+ b_{\tau i}(N\partial_j a_k - N\delta_j^x A_k^x - N\delta_j^y A_k^y - \partial_j a'_k)\epsilon^{ijk} + NB_\tau^x \delta_i^x \partial_j A_k^x \epsilon^{ijk} + NB_\tau^y \delta_i^y \partial_j A_k^y \epsilon^{ijk}]
\end{multline}
As in section \ref{subsec:4d2fol}, we integrate out $b_{\tau x}$, $b_{\tau y}$, $b_{\tau z}$, $A_\tau^x$, and $A_\tau^y$.  The constraints from integrating out $A_\tau^x$ and $A_\tau^y$ are addressed as before.  The novelty is the constraint from integrating out $b_{\tau z}$, which is
\begin{equation}
    \label{eq:hyb2folconstr1}
    N\partial_x a_y - N\partial_y a_x - \partial_x a'_y + \partial_y a'_x -N A_y^x +NA_x^y = 0.
\end{equation}
We incorporate this by defining
\begin{equation}
    \label{eq:hyb2foldef1}
    N\Tilde{A}_{xy} = NA_y^x + N\partial_y a_x - \partial_y a'_x,
\end{equation}
and demanding that it be symmetric in its indices.  We treat the constraints from integrating out $b_{\tau x}$ and $b_{\tau y}$ as in section \ref{subsec:4d2fol}.  Moreover, we define
\begin{equation}
    \label{eq:hyb2foldef2}
    \hat{A}_\tau^{xy} =  B_\tau^x - B_\tau^y
\end{equation}
\begin{equation}
    \label{eq:hyb2foldef3}
    \hat{A} = \partial_x B_y^x - \partial_y B_x^y - b_{xy}
\end{equation}
\begin{equation}
    \label{eq:hyb2foldef4}
    \hat{A}_z^{xy} = B_z^x - B_z^y,
\end{equation}
so that the fields have the gauge redundancy in \cref{eq:4d2folgt1,eq:4d2folgt2,eq:4d3folgt3} as before,  where the exotic gauge parameters are related to the foliated gauge parameters by \eqref{eq:4d2foldict}.  Written in terms of these fields, the Lagrangian is
\begin{multline}
    \label{eq:hyb2folcomp'}
    \mathcal{L} =  \frac{iN}{2\pi}[A_\tau(\partial_x \partial_y \hat{A}_z^{xy} - \partial_z \hat{A}) + \Tilde{A}_{xy}(\partial_\tau \hat{A}_z^{xy} - \partial_z \hat{A}_\tau^{xy}) + A_z (\partial_\tau \hat{A} - \partial_x \partial_y \hat{A}_\tau^{xy}) \\ + \frac{1}{2}B'_{\mu\nu}\partial_\rho A'_\sigma \epsilon^{\mu\nu\rho\sigma}] - \frac{i}{2\pi}[A'_\tau(\partial_x \partial_y \hat{A}_z^{xy} - \partial_z \hat{A}) + A'_z(\partial_\tau \hat{A} - \partial_x \partial_y \hat{A}_\tau^{xy})].
\end{multline}
This is not quite the desired result, as $\Tilde{A}_{xy}$ does not have the gauge redundancy of a symmetric hollow gauge field.  To rectify this, define
\begin{equation}
    \label{eq:hybrid2folredef}
    NA_{xy} = N\Tilde{A}_{xy} + \frac{1}{2}(\partial_x a'_y + \partial_y a'_x)
\end{equation}
in terms of which the Lagrangian is
\begin{multline}
    \label{eq:hybanis}
    \mathcal{L} = \frac{iN}{2\pi}[A_\tau(\partial_x \partial_y \hat{A}_z^{xy} - \partial_z \hat{A}) + A_{xy}(\partial_\tau \hat{A}_z^{xy} - \partial_z \hat{A}_\tau^{xy}) + A_z (\partial_\tau \hat{A} - \partial_x \partial_y \hat{A}_\tau^{xy}) \\ + \frac{1}{2}B'_{\mu\nu}\partial_\rho A'_\sigma \epsilon^{\mu\nu\rho\sigma}] - \frac{i}{2\pi}[A'_\tau(\partial_x \partial_y \hat{A}_z^{xy} - \partial_z \hat{A}) + A'_z(\partial_\tau \hat{A} - \partial_x \partial_y \hat{A}_\tau^{xy}) \\ + \frac{1}{2}(\partial_x A'_y + \partial_y A'_x)(\partial_\tau \hat{A}_z^{xy} - \partial_z \hat{A}_\tau^{xy})]
\end{multline}
The following periods are quantized:
\begin{equation}
    \oint (d\tau A_\tau +dz A_z) \in \frac{2\pi}{N^2}\mathbb{Z}
\end{equation}
\begin{equation}
    \oint (d\tau \hat{A}_\tau^{xy} + dz \hat{A}_z^{xy}) \in \frac{2\pi}{N}\mathbb{Z}
\end{equation}
\begin{equation}
    \int_{x_1}^{x_2}dx\oint (d\tau \partial_x A_\tau + dy A_{xy}) \in \frac{2\pi}{N^2}\mathbb{Z}
\end{equation}
\begin{equation}
    \int_{y_1}^{y_2}dy \oint (d\tau \partial_y A_\tau + dx A_{xy}) \in \frac{2\pi}{N^2}\mathbb{Z}
\end{equation}
\begin{equation}
    \int_{x_1}^{x_2}dx\oint (d\tau \partial_x \hat{A}_\tau^{xy} + dy \hat{A}) \in \frac{2\pi}{N}\mathbb{Z}
\end{equation}
\begin{equation}
    \int_{y_1}^{y_2}dy\oint (d\tau \partial_y \hat{A}_\tau^{xy} + dx \hat{A}) \in \frac{2\pi}{N}\mathbb{Z}
\end{equation}
\begin{equation}
    \oint_{\mathcal{C}^{(1)}} A' \in \frac{2\pi}{N}\mathbb{Z}
\end{equation}
\begin{equation}
    \oint_{\mathcal{C}^{(2)}} B' \in \frac{2\pi}{N^2}\mathbb{Z}
\end{equation}
This theory has the same operators discussed in the section \ref{subsec:4d2fol} are more or less present here, with the qualifier stemming from the fact that the quantized periods now imply:
\begin{multline}
    W_m^{N^2} = W_e^N = W_m(x,y)^{N^2} =W_e(x,y)^N = W_m(x_1,x_2)^{N^2} = W_m(y_1,y_2)^{N^2} \\ = W_e(x_1,x_2)^N =W_m(y_1,y_2)^{N^2}=1.
\end{multline}
Thus, these operators and defects are the exotic counterparts of those in the foliated field theory.  Moreover, it has the one and two form symmetries discussed in \cref{eq:1form,eq:2form,eq:12anomaly}.  Just as before, this provides three more N dimensional spaces to the calculation of the ground state degeneracy on $T^3$, so that we have 
\begin{equation}
    \label{eq:hybanisgsd}
    GSD = N^{2L_x+2L_y+1}.
\end{equation}
The second set of terms do not contribute to the ground state degeneracy.  What they \textit{do} contribute are the fusion rules that map N magnetic lineons to a mobile electric particle and N loops to an electric lineon.  As an example, we demonstrate the former.  The $\hat{A}^z$ equation of motion is 
\begin{equation}
    \label{eq:anishyb1}
    N (\partial_z A_\tau - \partial_\tau A_z) = \partial_z A'_\tau - \partial_\tau A'_z
\end{equation}
Now, consider fusing $N$ lineons:
\begin{multline}
    \label{eq:anishyb2}
    (W_m^z)^N = \exp[iN\oint_{\mathcal{C}^{\tau z}} (d\tau A_\tau + dz A_z)] = \exp[iN\int_{\mathcal{D}^{\tau z}} (d' (d\tau A_\tau + dz A_z))] \\ = \exp[i \int_{\mathcal{D}^{(2)}}(d'A')] = \exp[i\oint_{\mathcal{C}^{(1)}} A'] = W_e^{(1)},
\end{multline}
so fusing $N$ lineons gives a mobile particle.  Here, $\mathcal{C}^{\tau z}$ is a 1-cycle in $\tau z$ plane, $\mathcal{D}^{\tau z}$ is a surface diffeomorphic to a disk in the $\tau z$ plane such that $\partial \mathcal{D}^{\tau z}  = \mathcal{C}^{\tau z}$, and $d'$ is the exterior derivative in the $\tau z$ plane.  $\mathcal{C}^{(1)}$ and $\mathcal{D}^{(2)}$ are as defined below \cref{eq:1form,eq:tchyb2}.  The second and fourth equalities are from Stokes' theorem and the third equality is from equation \eqref{eq:anishyb2}.   The upshot is that fusing $N$ magnetic lineons gives a mobile electric particle.


\subsection{Three Foliations: Fractonic Hybrid X-Cube Model}
\label{subsec:hyb3fol}

Let $e^1 =dx=dx^1$, $e^2 = dy=dx^2$, and $e^3 = dz=dx^3$.  The Lagrangian is
\begin{multline}
    \label{eq:hyb3folcomp}
    \mathcal{L} = \frac{i}{2\pi}[\sum_{A=1}^3 N\delta_i^A B_j^A \partial_\tau A_k^A \epsilon^{ijk} + N\frac{1}{2}b_{ij}\partial_\tau a_k \epsilon^{ijk} + N\frac{1}{2}b'_{\mu\nu}\partial_\rho a'_\sigma \epsilon^{\mu\nu\rho\sigma} - \frac{1}{2}b_{ij}\partial_\tau a'_k \\+ N\sum_{A=1}^3A_\tau^A \delta_i^A(-\partial_j B_k^A + \frac{1}{2}b_{jk})\epsilon^{ijk} + (N\frac{1}{2}a_\tau \partial_i b_{jk} - \frac{1}{2}a'_\tau \partial_i b_{jk})\epsilon^{ijk} \\+ b_{\tau i}(N\partial_j a_k - N\sum_{A=1}^3\delta_j^A A_k^A - \partial_j a'_k)\epsilon^{ijk} + N\sum_{A=1}^3B_\tau^A \delta_i^A \partial_j A_k^A \epsilon^{ijk}]
\end{multline}
As in section \ref{subsec:4d3fol}, we integrate out $b_{\tau i}$ and $A_\tau^i$ for all $i$.  The constraints from integrating out $b_{\tau i}$ are treated as in section \ref{subsec:4d3fol}, whereas we treat the constraints from integrating out $A_\tau^i$ as in section \ref{subsec:hyb2fol}.  That is, we define
\begin{equation}
    \label{eq:hyb3foldef1}
    N\Tilde{A}_{ij} = NA_j^i + N\partial_j a_i - \partial_j a'_i.
\end{equation}
Moreover, as in section \ref{subsec:4d3fol}, we define:
\begin{equation}
    \label{eq:hyb3foldef2}
    \hat{A}_\tau^{k(ij)} = B_\tau^i - B_\tau^j
\end{equation}
\begin{equation}
    \label{eq:hyb3foldef3}
    \hat{A}^{ij} = B_k^i - B_k^j
\end{equation}
These gauge fields have the gauge redundancy in \cref{eq:4d3folgt1,eq:4d3folgt2,eq:4d3folgt3} with the exotic gauge parameters related to the foliated gauge parameters as in equation \eqref{eq:4d3foldict}.  In terms of these, the Lagrangian is 
\begin{equation}
    \label{eq:hyb3folcomp'}
    \mathcal{L} = \frac{iN}{4\pi}[\Tilde{A}_{ij}(\partial_\tau \hat{A}^{ij} - \partial_k \hat{A}_\tau^{k(ij)})+A_\tau (\partial_i\partial_j\hat{A}^{ij}) + \epsilon^{\mu\nu\rho\sigma}B'_{\mu\nu}\partial_\rho A'_\sigma]-\frac{i}{4\pi}[A'_\tau \partial_i \partial_j \hat{A}^{ij}].
\end{equation}
Of course, just as in section \ref{subsec:hyb2fol}, we are note quite done.  To rewrite the theory in terms of the desired gauge fields, we need to define
\begin{equation}
    \label{eq:hyb3folredef}
    NA_{ij} = N\Tilde{A}_{ij} + \partial_i a'_j + \partial_j a'_i,
\end{equation}
so that the Lagrangian takes the form
\begin{multline}
    \label{eq:hybXcube}
    \mathcal{L} = \frac{iN}{4\pi}[A_{ij}(\partial_\tau \hat{A}^{ij} - \partial_k \hat{A}_\tau^{k(ij)})+A_\tau (\partial_i\partial_j\hat{A}^{ij}) + \epsilon^{\mu\nu\rho\sigma}B'_{\mu\nu}\partial_\rho A'_\sigma] \\ -\frac{i}{4\pi}[\partial_i A'_j (\partial_\tau \hat{A}^{ij} - \partial_k \hat{A}_\tau^{k(ij)}) +A'_\tau (\partial_i \partial_j \hat{A}^{ij})]
\end{multline}
One can show the following periods to be quantized:
\begin{equation}
    \oint d\tau A_\tau \in \frac{2\pi}{N^2}\mathbb{Z}
\end{equation}
\begin{equation}
    \int_{x_1}^{x_2} dx \oint (d\tau \partial_x A_\tau + dy A_{xy} + dz A_{zx}) \in \frac{2\pi}{N^2}\mathbb{Z} 
\end{equation}
\begin{equation}
    \int_{y_1}^{y_2} dy \oint (d\tau \partial_y A_\tau + dx A_{xy} + dz A_{yz}) \in \frac{2\pi}{N^2}\mathbb{Z} 
\end{equation}
\begin{equation}
    \int_{z_1}^{z_2} dz \oint (d\tau \partial_z A_\tau + dx A_{zx} + dy A_{yz}) \in \frac{2\pi}{N^2}\mathbb{Z} 
\end{equation}
\begin{equation}
    \oint (d\tau \hat{A}_\tau^{k(ij)} + dx^k \hat{A}^{ij}) \in \frac{2\pi}{N}\mathbb{Z}
\end{equation}
\begin{equation}
    \oint_{\mathcal{C}^{(1)}} A' \in \frac{2\pi}{N}\mathbb{Z}
\end{equation}
\begin{equation}
    \oint_{\mathcal{C}^{(2)}} B' \in \frac{2\pi}{N^2}\mathbb{Z}
\end{equation}
The defects and operators discussed in the section \ref{subsec:4d3fol} are still present here.  However, since the periods are quantized differently, they fuse differently.  In particular, we have:
\begin{multline}
    W_m(x,y,z)^{N^2} = W_m(x_1,x_2,\mathcal{C}^{yz})^{N^2} = W_m(y_1,y_2,\mathcal{C}^{zx})^{N^2} = W_m(z_1,z_2,\mathcal{C}^{xy})^{N^2} \\= \hat{W}_e^x(y,z)^N = \hat{W}_e^y(x,z)^N = \hat{W}_e^z(x,y)^N = 1.
\end{multline}
Thus, these operators/defects are the exotic counterparts to the holonomies in the foliated field theory.  Moreover, it has the one and two form symmetries discussed in \cref{eq:1form,eq:2form,eq:12anomaly}.  Just as before, this provides three more N dimensional spaces to the calculation of the ground state degeneracy on $T^3$, so that we have
\begin{equation}
    \label{eq:hybXcubegsd}
    GSD = N^{2L_x+2L_y+2L_z}
\end{equation}
The second set of terms do not contribute to the ground state degeneracy.  What they \textit{do} contribute are fusion rules mapping N fractons to a mobile particle and N loops to a lineon.  As an example, we demonstrate the former.  Since planons are fracton dipoles, we work with those, expecting a dipole of mobile particles.  The $\hat{A}_\tau^{k(ij)}$ equation of motion is
\begin{equation}
    \label{eq:Xcubehyb1}
    \partial_k \partial_i A'_j = N\partial_k A_{ij}
\end{equation}
Now, imagine fusing $N$ planons (in the x-y plane for definiteness):
\begin{multline}
    \label{eq:Xcubehyb2}
    W(z_1,z_2)^N = \exp[iN\int_{z_1}^{z_2}dz\oint_{\mathcal{C}^{xy}}(dx A_{zx} + dy A_{yz})] \\ = \exp[iN\int_{z_1}^{z_2}dz \int_{\mathcal{D}^{xy}} dx dy(\partial_y A_{zx} + \partial_x A_{yz})] = \exp[i\int_{z_1}^{z_2}dz\int_{\mathcal{D}^{(2)}} dx dy(\partial_y \partial_z A'_x + \partial_x \partial_z A'_y)]\\ = \exp[i\int_{z_1}^{z_2}dz \partial_z \int_{\mathcal{C}^{(1)}} (dx A'_x + dy A'_y)],
\end{multline}
where $\mathcal{C}^{xy}$, $\mathcal{D}^{xy}$, $\mathcal{D}^{(2)}$, and $\mathcal{C}^{(1)}$ are defined below equations \cref{eq:1form,eq:tchyb2}.  The second and fourth equalities follow from Stokes' theorem and the third equality follows from the above equation of motion.  We see that fusing $N$ planons results in a dipole of mobile particles.   

Therefore, this simple field theory, which we have not seen in this form in the literature, encapsulates the key physics of the fractonic hybrid X-Cube introduced in \cite{Hybrid}.

\section{Conclusion and Discussion}
\label{sec:conc}

In this paper, we introduced a simple recipe for extracting exotic field theories from foliated field theories - integrate out the Lagrange multipliers that relate foliated gauge fields to the standard gauge fields.  We applied this recipe to a variety of foliated field theories, uniting a mixture of old and new results.  We began with the $2+1$ dimensional foliated BF theory and map it to the $2+1$ dimensional exotic BF theory, rediscovering the map between gauge fields in \cite{FoliatedExotic}.  We then moved to the $3+1$ dimensional foliated BF theory, which we analyzed for 1, 2, and 3 flat foliations.  For one foliation we obtain an exotic field theory for the toric code layers that we have not encountered previously in the literature.  For two foliations, we obtain the exotic theory for the anisotropic theory with lineons and planons.  For three foliations, we obtain the exotic BF theory for the X-Cube model and recreate the map between operators in \cite{FoliatedExotic}.  In all three cases, our map is the natural continuum version of the work done on the lattice in \cite{SAW}.  Our next targets were foliated field theories shown in \cite{ElectricMagneticModels} to contain hybrid fracton phenomenology.  Our procedure supplies new exotic field theories for the hybrid phases discussed in \cite{Hybrid} that give a transparent way to account for hybrid fracton behavior in exotic field theory that begs to be generalized by future work.     

Our results lead to a plethora of directions for further work:
\begin{itemize}
    \item Throughout this paper, we examined a number of foliations less than or equal to the number of spatial dimensions.  It would be interesting to apply the construction herein to theories with more foliations than spatial directions.  In particular,  Chamon's model \cite{Chamon} is a four foliated fracton phase \cite{Chamon2} that is related to the four foliated X-Cube.  Examining four-foliated theories and comparing the result to the field theories discussed in \cite{FractonicCS,ChamonQFT} would be fruitful.
    \item The foliations discussed in this paper are all along the x,y, or z direction.  It would be interesting to discuss foliations with curved leaves, or foliations with nontrivial Godbillon-Vey invariant.
    \item In addition to the magnetic models with a closed two form $b$, \cite{ElectricMagneticModels} discusses electric models with a closed one form $a$.  Obtaining exotic field theories for electric models would expand the scope of foliated field theories from which wee know how to obtain an exotic field theory.  \cite{ElectricMagneticModels} uncovers a duality between $G$ electric models and $G/\mathbb{Z}_N$ magnetic models.  It would be fruitful to see how this duality appears in exotic field theory. 
    \item In \cite{DefectNet}, the authors capture the mobility constraints of the X-Cube by embedding a network of condensation defects in the three dimensional toric code.  It might be illustrative to connect that viewpoint to quantum field theory, particularly since defect networks can also encapsulate the mobility constraints of type II fractons. 
    \item The map from foliated to exotic field theory presented in the paper is only true if integrating by parts simply transports partial derivatives and gives minus signs.  Generally, this is not true if the manifold on which the field theory lives is not closed.  Thus, one should ask which boundary conditions are suitable for the duality in the paper.  This could be particularly subtle in light of the fact that BF theories are not gauge invariant on manifolds with a boundary \cite{CMQFT,XCubeBndy,FractonicCS}.
    \item We do not discuss gapless theories at all in this paper, despite them appearing in both foliated \cite{ElectricMagneticModels} and exotic \cite{SS1,SS2} settings.  Understanding the relationship between gapless foliated and gapless exotic theories is an out standing problem.
    \item Our recipe for hybrid fracton phenomenology is straightforward and generalizable - simply couple the exotic theory and the BF theory to a theory that doesn't contribute to the ground state degeneracy and enforces the appropriate fusion rules as equations of motion.  It would be intriguing to see what hybrid phases one can write down this way that do not involve the X-Cube.  Also, \cite{ElectricMagneticModels} contains more complicated magnetic models whose exotic field theory remains to be uncovered.  Some of these correspond to non-Abelian hybrid fracton phases, introduced in \cite{NonAbelianHybrid}.  Moreover, coupling exotic field theories to more familiar field theories seems to be a frontier ripe for exploration.  
\end{itemize}

\acknowledgments

We are grateful to Zhu-Xi Luo for referring us to \cite{SAW} and to Ho Tat Lam for a clarification on the anisotropic model.  We thank Andreas Karch and Kevin Slagle for comments on the draft and related conversations.  We also thank an anonymous JHEP referee for helpful input, especially for helpful comments about the anisotropic theory and for encouraging us to discuss the periods of the gauge fields in some detail.

This work was supported, in part, by the U.S.~Department of Energy under
Grant DE-SC0022021 and by a grant from the Simons Foundation (Grant
651440, AK).

\appendix
\section{Quantized Periods in the Fractonic Hybrid X-Cube}
In this appendix, we demonstrate the quantized periods in both the foliated and exotic presentation of the Fractonic Hybrid X-Cube.  The periods in the other theories can be derived analogously.  Our discussion was informed by \cite{FoliatedQFT,FoliatedExotic}.  We will find that, although we begin with $U(1)$ gauge fields with the standard normalization, the BF structure of the actions imposes additional constraints that turn them into finite group gauge fields.  

We begin on the foliated side.  For convenience, the action is
\begin{equation}
    \label{eq:hybfol'}
    \mathcal{L} = \frac{iN}{2\pi}[\sum_{A=1}^3 (dx^A \wedge B^A \wedge dA^A - dx^A\wedge b\wedge A^A) + b\wedge da + b'\wedge da'] - \frac{i}{2\pi}b\wedge da'. 
\end{equation}
We consider the theory on an untwisted spacetime four torus with lengths $l^\tau$, $l^x$, $l^y$, and $l^z$.  Let us first address the standard gauge field $a'$.  Consider the configuration
\begin{equation}
    b' = -\frac{1}{2}j d(cos\theta)\wedge d\phi ; j\in \mathbb{Z}
\end{equation}
such that
\begin{equation}
    db' = 2\pi j \delta^3(x) dx\wedge dy\wedge dz.
\end{equation}
The term in the action involving this configuration is
\begin{equation}
    -iNj\oint_{\mathcal{C}^\tau} a' 
\end{equation}
To integrate out $b'$, sum over $j$.  This implies: 
\begin{equation}
    \oint_{\mathcal{C}^\tau} a' \in \frac{2\pi}{N}\mathbb{Z}.
\end{equation}
Using the equation of motion $da'=0$, we can deform $\mathcal{C}^\tau$ within its homology class, giving
\begin{equation}
    \oint_{\mathcal{C}^{(1)}}a' \in \frac{2\pi}{N}\mathbb{Z},
\end{equation}
since we can obviously proceed similarly in other directions.
We now move to $b$.  Consider
\begin{equation}
    a = \frac{1}{2}j\frac{2\pi}{l^x l^y}(-ydx + x dy); j \in \mathbb{Z}
\end{equation}
such that
\begin{equation}
    da = j\frac{2\pi}{l^x l^y} dx\wedge dy.
\end{equation}
The term in the action containing this configuration is
\begin{equation}
    iNj\oint_{\mathcal{C}^{\tau z}} b.
\end{equation}
Summing over $j$ to integrate out $a$ gives
\begin{equation}
    \oint_{\mathcal{C}^{\tau z}} b \in \frac{2\pi}{N}\mathbb{Z}.
\end{equation}
Using the equation of motion $db = 0$, we can deform $\mathcal{C}^{\tau z}$ in its homology class.  Moreover, since we can obviously work analogously in other directions, we have shown
\begin{equation}
    \oint_{\mathcal{C}^{(2)}} b \in \frac{2\pi}{N}\mathbb{Z}.
\end{equation}
Using the equation of motion $b = Nb'$, we obtain 
\begin{equation}
    \oint_{\mathcal{C}^{(2)}} b' \in \frac{2\pi}{N^2}\mathbb{Z}.
\end{equation}
Now, consider configurations such that
\begin{equation}
    dA^y = j \frac{2\pi}{l^xl^yl^z}dx\wedge dy\wedge dz; j \in \mathbb{Z} 
\end{equation}
and
\begin{equation}
    dA^z = -j \frac{2\pi}{l^xl^yl^z}dx\wedge dy\wedge dz; j \in \mathbb{Z}.
\end{equation}
As discussed in \cite{FoliatedQFT}, one can choose $a$ such that $da + \sum_A A^A \wedge dx^A = 0$ so that the term in the action with these configurations is  
\begin{equation}
    ijN\oint_{\mathcal{C}_1^\tau}(B^y-B^z).
\end{equation}
Upon summing over $j$ to integrate out $A^y$ and $A^z$, we obtain
\begin{equation}
    \oint_{\mathcal{C}_1^\tau}(B^y-B^z) \in \frac{2\pi}{N}\mathbb{Z}.
\end{equation}
Using the equations of motion $(dB^y+b)\wedge dy =0$ and $(dB^z+b)\wedge dz=0$, we can deform $C_1^\tau$ in the $\tau-x$ plane.  Thus, we have found
\begin{equation}
    \oint_{\mathcal{C}_1^{\tau x}}(B^y-B^z) \in \frac{2\pi}{N}\mathbb{Z}.
\end{equation}
Similar arguments give similar results in the other directions. Now, consider 
\begin{equation}
    b = -\frac{1}{2}j d(cos\theta)\wedge d\phi ; j \in \mathbb{Z}
\end{equation}
such that
\begin{equation}
    db = j2\pi \delta^3(x) dx\wedge dy \wedge dz.
\end{equation}
To ensure that the equation of motion $(dB^A + b)\wedge dx^A = 0$ is satisfied, we pick
\begin{equation}
    B^A = \frac{1}{2}j(cos\theta -1)d\phi 
\end{equation}
so that the term in the action containing the configurations we analyze is
\begin{equation}
    -iNj \oint_{\mathcal{C}^\tau} a + ij \oint_{\mathcal{C}^\tau}a' 
\end{equation}
Summing over $j$ gives
\begin{equation}
    \oint_{\mathcal{C}^\tau}(-Na +a') \in 2\pi \mathbb{Z},
\end{equation}
which, upon invoking the quantization of $a'$, gives
\begin{equation}
    \oint_{\mathcal{C}^\tau}a \in \frac{2\pi}{N^2}\mathbb{Z}. 
\end{equation}
Now, we turn to $A^A \wedge dx^A$.  We consider the z direction (as usual, the others can be addressed similarly) and analyze $B^z$ such that
\begin{equation}
    dB^z = j\frac{2\pi}{l^x l^y} dx\wedge dy ; j \in \mathbb{Z}.
\end{equation}
The term in the action with the configuration is
\begin{equation}
    ijN\oint_{\mathcal{M}_2^P}A^z \wedge dz 
\end{equation}
so that summing over $j$ gives
\begin{equation}
    \oint_{S^z}A^z \wedge dz \in \frac{2\pi}{N}\mathbb{Z}.
\end{equation}
Using the equation of motion $d(A^z\wedge dz)=0$, we can deform this on leaves of the foliation defined by $dz$.  This is not what we match across the duality.  Instead, we match terms such as $A^x \wedge dx + d(a_x dx)$.  To address this, note that integrating out $a'$ gives the theory \footnote{Note that this theory has the quantized periods addressed so far (obviously excluding the field that was integrated out).}:
\begin{equation}
    \mathcal{L} = \frac{iN}{2\pi}\sum_{A=1}^3 dx^A\wedge B^a \wedge dA^A - \frac{iN^2}{2\pi}(\sum_{A=1}^3 dx^A \wedge b' \wedge A^A + b'\wedge da).
\end{equation}
Now, consider
\begin{equation}
    b' = 2\pi j \frac{1}{l^y l^z}(\Theta(x-x_1) - \Theta(x-x_2))dy\wedge dz; j \in \mathbb{Z}.
\end{equation}
The term in the action containing this configuration is
\begin{equation}
    ij \oint_{\mathcal{C}^\tau \times [x_1,x_2]} [N^2(A_\tau^x + \partial_\tau a_x - \partial_x a_\tau)]
\end{equation}
Summing over $j$ implies
\begin{equation}
    \oint_{\mathcal{C}^\tau \times [x_1,x_2]} [N^2(A_\tau^x + \partial_\tau a_x - \partial_x a_\tau)] \in 2\pi \mathbb{Z}
\end{equation}
Using the known quantization of $a_\tau$, this reduces to
\begin{equation}
    \oint_{\mathcal{C}^\tau \times [x_1,x_2]}[(A_\tau^x + \partial_\tau a_x)]  \in \frac{2\pi}{N^2}\mathbb{Z}.
\end{equation}
Using the equation of motion $d(A^x\wedge dx) = 0$, we can deform on leaves of the $x$ foliation, so we have obtained
\begin{equation}
    \oint_{\mathcal{C}^{(1)} \times [x_1,x_2]}[(A^x \wedge dx + d(a_x dx))]  \in \frac{2\pi}{N^2}\mathbb{Z}.
\end{equation}

We now turn to the exotic presentation of the theory.  For convenience, the Lagrangian is:
\begin{multline}
    \label{eq:hybXcube'}
    \mathcal{L} = \frac{iN}{4\pi}[A_{ij}(\partial_\tau \hat{A}^{ij} - \partial_k \hat{A}_\tau^{k(ij)})+A_\tau (\partial_i\partial_j\hat{A}^{ij}) + \epsilon^{\mu\nu\rho\sigma}B'_{\mu\nu}\partial_\rho A'_\sigma] \\ -\frac{i}{4\pi}[\partial_i A'_j (\partial_\tau \hat{A}^{ij} - \partial_k \hat{A}_\tau^{k(ij)}) +A'_\tau (\partial_i \partial_j \hat{A}^{ij})].
\end{multline}
An identical argument to the one we just gave informs us that
\begin{equation}
    \oint_{\mathcal{C}^{(1)}} A' \in \frac{2\pi}{N}\mathbb{Z}.
\end{equation}
We now move to quantizing the periods of the hatted fields.  Consider:
\begin{equation}
    A_{xy} = 2\pi j \frac{\tau}{l^\tau}[\frac{1}{l^y}\delta(x-x_0) + \frac{1}{l^x}\delta(y-y_0) - \frac{1}{l^xl^y}] ; j \in \mathbb{Z}.
\end{equation}
Plugging into the action gives
\begin{equation}
    -ijN\oint dz \hat{A}^{xy} 
\end{equation}
so that summing over $j$ imposes the constraint
\begin{equation}
    \oint dz \hat{A}^{xy} \in \frac{2\pi}{N}\mathbb{Z}.
\end{equation}
Using the equation of motion $\partial_\tau \hat{A}^{xy} - \partial_z \hat{A}_\tau^{z(xy)}=0$, we can deform this in the $\tau-z$ plane, so we find:
\begin{equation}
    \oint_{\mathcal{C}_1^{\tau z}} (d\tau \hat{A}_\tau^{z(xy)} + dz \hat{A}^{xy}) \in \frac{2\pi}{N}\mathbb{Z}
\end{equation}
We can use similar arguments to get analogous results in other directions.  We now move to unhatted fields.  Consider:
\begin{equation}
    \hat{A}^{xy} = j\frac{2\pi}{l^y}y\partial_z[\frac{z}{l^z}\Theta(x-x_0)+\frac{x}{l^x}\Theta(z-z_0) - \frac{xz}{l^x l^z}]; j \in \mathbb{Z}.
\end{equation}
Plugging into the action gives
\begin{equation}
    ij \oint d\tau (NA_\tau - A'_\tau). 
\end{equation}
Summing over $j$ then gives the constraint
\begin{equation}
    \oint d\tau (NA_\tau - A'_\tau) \in 2\pi \mathbb{Z},
\end{equation}
which, upon using the quantization of $\oint d\tau A'_\tau$, gives
\begin{equation}
    \oint d\tau A_\tau \in \frac{2\pi}{N^2}\mathbb{Z}.
\end{equation}
We can use this to address other periods.  For instance, this immediately gives
\begin{equation}
    \int_{z_1}^{z_2}dz\oint d\tau \partial_x A_\tau \in \frac{2\pi}{N^2}\mathbb{Z},
\end{equation}
which, using the equations of motion $\partial_\tau A_{zx} - \partial_z \partial_x A_\tau + \frac{1}{2}(\partial_\tau \partial_y A'_z + \partial_\tau \partial_z A'_y) - \partial_y\partial_z A'_\tau=0$, $\partial_\tau A_{yz} - \partial_y \partial_z A_\tau + \frac{1}{2}(\partial_\tau \partial_z A'_x + \partial_\tau \partial_x A'_z) - \partial_z\partial_x A'_\tau =0$, $\partial_x A_{yz} - \partial_y A_{zx} + \frac{1}{2}\partial_z(\partial_y A'_x - \partial_x A'_y) =0$, and $\epsilon^{\mu\nu\rho\lambda} \partial_\rho A'_\lambda=0$, can be deformed into
\begin{equation}
    \int_{z_1}^{z_2}dz\oint(d\tau \partial_x A_\tau + dx A_{zx} + dy A_{yz})\in \frac{2\pi}{N^2}\mathbb{Z}.
\end{equation}
Identical arguments work for the other directions. 

We leave a direct argument for the quantization of $B'$'s periods to future work.  For an indirect argument, note that the duality dictionary is $B'=b'$, so one expects the quantization 
\begin{equation}
\oint_{\mathcal{C}^{(2)}}B' \in \frac{2\pi}{N^2}\mathbb{Z}. 
\end{equation}



\end{document}